%% file: paper.tex
\documentclass[reprint,
aps,
prd,
twocolumn,
showpacs,
superscriptaddress,
groupedaddress]{revtex4}

\usepackage{graphicx}
\usepackage{dcolumn}
\usepackage{bm}
\graphicspath{{./Figures/}}

\usepackage[dvipsnames]{xcolor}

\include{macroses}

\begin{document}

\title{
New way of collision experiment data analysis based on Grand Canonical Distribution and Lattice QCD data
}

\author{V. Bornyakov}
\affiliation{School of Biomedicine, Far Eastern Federal University, 690950 Vladivostok, Russia}

\author{D. Boyda}
\email{boyda\_d@mail.ru}
\affiliation{School of Biomedicine, Far Eastern Federal University, 690950 Vladivostok, Russia}

\author{V. Goy}
\affiliation{School of Biomedicine, Far Eastern Federal University, 690950 Vladivostok, Russia}

\author{A. Molochkov}
\affiliation{School of Biomedicine, Far Eastern Federal University, 690950 Vladivostok, Russia}

\author{A. Nakamura}
\affiliation{School of Biomedicine, Far Eastern Federal University, 690950 Vladivostok, Russia}

\date{\today}

\begin{abstract}

We propose new way of heavy ion collisions experiment data analysis. We analyze
physical parameters of fireball created in RHIC experiment based on Grand Canonical Distribution and different Lattice QCD data available at the moment. Our results on chemical potential are in agreement with previous model estimations and do not depend on Lattice setup. 
At same time, we found possible T(V) states of fireball and estimated the most probable temperature and volume of fireball as function of collision energy. 
We conclude that hadrom matter at RHIC experiment is thermalized and described by Grand Canonical Distribution.
\end{abstract}


\maketitle


\section{\label{sec:Intro}Introduction}
Analysis of experimental data is highly non trivial matter \cite{RHICReview2019}. Usually in heavy ion collisions multiplicity distributions as well as fluctuations are measured. Using these data and different models one can extract physical parameters of freeze-out such as temperature, chemical potential and volume. For example authors of \cite{Alba} calculated higher moments in a model and by comparing their results with experimental data found chemical potential and temperature of freeze-out conditions. It is more interesting to analyze experimental data with LQCD. Unfortunately because of Sign Problem progress in LQCD with non zero chemical potential is rather slow. Nevertheless, in paper \cite{Karsch2017} authors present precise data of Lattice QCD calculations in Taylor expansion approach (where LQCD simulations is conducted at zero chemical potential) and calculate QCD equation of state. They calculate line of constant physics (pressure, energy density and entropy density) for different initial temperatures and compare it with experimental data analysis in STAR and ALICE. Usually, in these studies one uses experimental moments like variance, kurtosis and skewness but it seems that multiplicity distribution have much information as well. Indeed, as it was shown in \cite{PTEP_Nakamura} one can extract canonical functions $Z_n$ from experimental multiplicities. Using them authors of \cite{2017Boyda} calculated observables and compared  it with Lattice QCD data.

In this paper we present new way of analyzing experimental data. In assumption that experimental states have canonical distribution we build connection of multiplicity and Canonical Functions $Z_n$  which can be calculated using LQCD simulation at imaginary chemical potential. Using two different sets of LQCD results we analyse temperature, chemical potential and volume of fireball. Our conclusions are in qualitative agreement with previous model analysis.

\section{Grand Canonical Distribution}

Let us start with Grand Canonical Distribution
\beq
Z_{GC}(\mu, T, V) = \sum_n Z_n(T, V) e^{n\mu/T},
\eeq
which is used to represent the possible states of a statistical system of particles that are in thermodynamic equilibrium (thermal and chemical) with a reservoir. 
The thermodynamic variables of this ensemble are chemical potential $\mu$, temperature $T$ and volume $V$. The net baryon number fluctuates. From this equation one can easily get normalization condition 
\beq
\sum_n \frac{Z_n(T, V) e^{n\mu/T}}{Z_{GC}(\mu, T, V)} = 1
\eeq
for probabilities of states with fixed $n, \mu, T, V$. Therefore to calculation number of states with fixed $n$,  $T$, $V$ and $\mu$ created after freeze-out we can use relation
\beq \label{eq:PnDefWithProb}
P(n) = N \frac{Z_n(T, V) e^{n\mu/T}}{Z_{GC}(\mu, T, V)},
\eeq
with total number of events $N$. Quantity $P(n)$ is nothing else but multiplicity measured in experiment. Unfortunately, baryon multiplicity is never measured in experiments due to complexity of identification baryon states. Instead, collision experiments usually provide us net proton multiplicity $P_n$. In this paper we use common assumption that baryon distribution is similar to proton one. There are some arguments for numerical similarity between  the net-proton and net-baryon number multiplicity distributions. For details see, e.g. \cite{Ratti:2019tvj}. It is easy to obtain from (\ref{eq:PnDefWithProb})
the relation \cite{PTEP_Nakamura}:
\beq\label{eq:PnFormula}
\frac{P_n}{P_0}(\mu, T, V) = \frac{Z_n}{Z_0}(T, V) e^{n\mu/T},
\eeq
which we will use in our analysis.

Canonical functions $Z_n$ can be calculated via Fourier transformation of $Z_{GC}$ computed at imaginary baryon chemical potential \cite{Roberge:1986mm,hasenfratz1992canonical}. 
Recently \cite{CA2016PRD} we computed $Z_n$ in $N_f=2$ LQCD using Integration Method:
\beq\label{eq:ZnGeneralFormula}
\frac{Z_n}{Z_0} (T, V)  = \frac{\int_0^{2\pi}d\theta e^{in\theta}e^{-V\int_0^{\theta} dx \, n^{Lat}(T, x) } }
{\int_0^{2\pi}d\theta e^{-V\int_0^{\theta} dx \, n^{Lat}(T, x) }},
\eeq
where $i\,n^{Lat}(T, x)$ is imaginary number density computed on the lattice at imaginary chemical $x=\mu_{Im}/T$.

It is worth to note that number density in general is local quantity but in the lattice calculations one calculates average quantity and volume dependence may exist if system is not homogeneous. Nevertheless we conducted LQCD simulations for different volumes and found no volume dependence. Therefore $Z_n$ depends on volume only due to the factor  $V=(aN_s)^3$ in Eq. \ref{eq:ZnGeneralFormula}, where $a$ is lattice spacing, $N_s$ is number of lattice sites in spatial direction. In LQCD calculation variation of temperature $T=1/(aN_t)$, where $N_t$ is number of lattice sites in time direction, implies changing lattice spacing $a$ and, therefore, volume. Thus it is useful to normalize $Z_n$ at all temperatures to same volume. It is quite convenient to choose volume as $V=V_c = (N_sa_c)$:
\beq
V_cn^{Lat} = N_s^3 a_c^3n^{Lat} = N_s^3 (T/T_c)^3 a^3 n^{Lat},
\eeq
where $T_c$ is the temperature of the confinement-deconfinement transition, $a_c$ is
respective lattice spacing. Numerical calculation of number density integral with sufficient precision requires LQCD simulations at many values of $\mu$ within interval $[0,\pi/3]$ which is quite expensive in terms of computer time. It was shown many times in literature \cite{CA2016PRD,2004Lombardo, Delia2009,takahashi2015quark} that number density can be well approximated by few terms of the Fourier sine series at $T<T_{RW}$ where $T_{RW}$ is the so called Roberge-Weiss temperature which value is somewhat above $T_c$:
\beq\label{eq:numDens}
n^{Lat}(\mu/T) = \sum_{k=1}^{K} f_k^{Lat}(T) \, sin(3k\mu/T).
\eeq
Finally, for calculation of multiplicities from Lattice data in Confinement phase one comes to the formula
\beq\label{eq:PnFinalFormula}
\frac{P_n}{P_0} (\mu, T, V= \kappa V_c) =  
e^{n\frac{\mu}{T}}
 \frac{\int d\theta e^{in\theta}
 	e^{\kappa \sum_{k=1}^{K} \bar{f_k}^{Lat}(T) \, cos(3k\theta) } }
{\int d\theta 
	e^{\kappa \sum_{k=1}^{K} \bar{f_k}^{Lat}(T) \, cos(3k\theta) }
},
\eeq
with coefficients $\bar{f_k} = \frac{a^3f_k^{Lat}(T)}{3k}N_s^3(T/T_c)^3$ extracted from Lattice QCD.
It is worth to note that in opposite to volume one can not vary temperature continuously. Dependence of $P_n(V)$ on volume implies changing $\kappa$ in Eq. \ref{eq:ZnGeneralFormula} while changing of temperature means running LQCD simulations which is quite expensive. Therefore we applied cubic spline interpolation using as input lattice results computed at few values of temperature. 

\begin{figure}[h]
	\includegraphics[width=\linewidth]{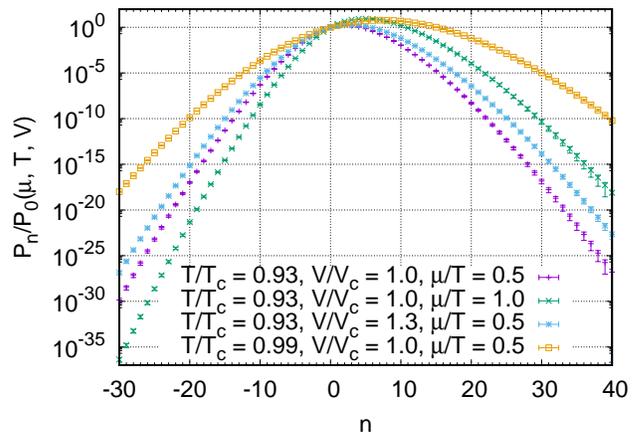}
	\caption{
		Multiplicity calculated from LQCD data \cite{CA2016PRD} (see details in the text) for different parameters: temperature, volume and chemical potential.
	}
	\label{fig:PnLat}
\end{figure}

To study $\frac{P_n}{P_0} (\mu, T, V)$ dependence on parameters we used Lattice data published in \cite{CA2016PRD} (Tab. II) where simulations were done with clover improved Wilson and Iwasaki gauge actions on a lattice $4\times16^3$ for pion mass equal approximately 700 MeV. Results of calculation of multiplicity for different values of temperature, volume and chemical potential is presented on Fig. \ref{fig:PnLat}.
Dependence of multiplicity on temperature agrees with Boltzmann statistics. 
The fact that multiplicity at fixed $n$ increases with temperature can be explained in such a way that probability of state $n$ with fixed energy increases with temperature. Nevertheless we stress that $P_n/P_0$ is not probability but only quantity proportional to it. Also from Fig. \ref{fig:PnLat} we see that the higher $n$ the stronger temperature dependence. Volume dependence of multiplicity is similar to dependence on temperature. In opposite, increasing of the chemical potential results in rise of multiplicity for large values of $n$ which can be explained by increasing asymmetry of baryons.

\hfill

\section{Analysis of RHIC data with Grand Canonical Distribution}

In this paper we analyze RHIC experiment data \cite{RHICdata1,RHICdata2}. Our goal is to check how well experimental data can be approximated by Eq. \ref{eq:PnFinalFormula} with different LQCD data or, in other words, if experimental data are distributed with Grand Canonical Distribution. We proceed with chi-squared fitting of experimental multiplicities at different energies to Eq. \ref{eq:PnFinalFormula} with T, V, $\mu$ as parameters.

\begin{figure}[h]
	\includegraphics[width=\linewidth]{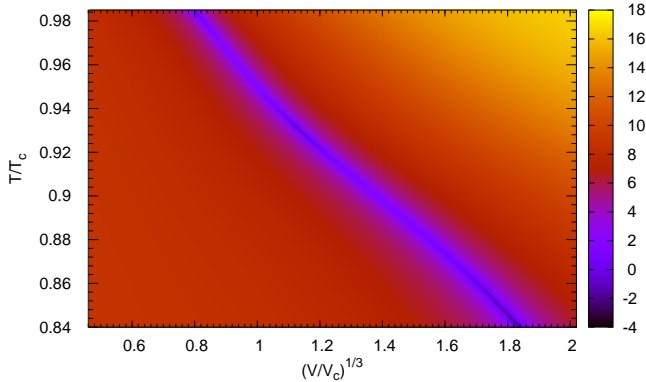}
	\caption{
		Logarithm of chi-squared value for fit of the RHIC experimental data at 
		200 GeV to multiplicity calculated with 
		Eq. \ref{eq:PnFinalFormula}. LQCD data from \cite{CA2016PRD} are used. The value is indicated by color. The chemical potential is fixed to $\mu/T=0.16$, $V_c \approx (4.5\,\text{fm})^3$.
	}
	\label{fig:cmp4_chisq200}
\end{figure}
\begin{figure}[h]
	\includegraphics[width=\linewidth]{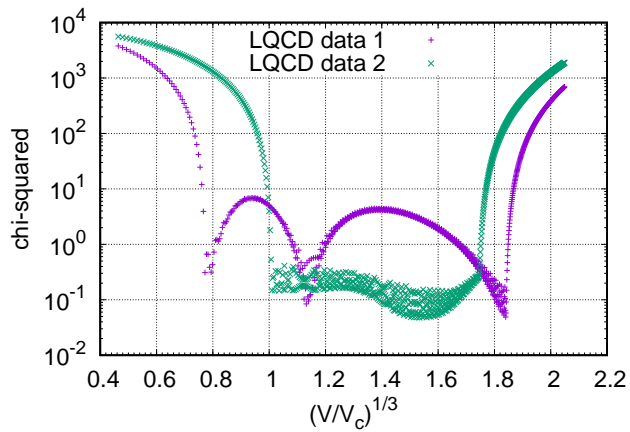}
	\caption{
		Chi-squared value along the line $T=T(V)$ of its minima
		visualized in Fig. \ref{fig:cmp4_chisq200} by dark blue color. 
		LQCD data from \cite{CA2016PRD} (LQCD data 1, $V_c \approx (4.5\,\text{fm})^3$) and \cite{Fodor2017} (LQCD data 2, $V_c \approx (5.1\,\text{fm})^3$) are used. For \cite{CA2016PRD}
	}
	\label{fig:chisq_valley200}
\end{figure}

From Eq. \ref{eq:PnFormula} one see that dependence of $P_n$ on $\mu$ and on $T,V$ factories so determination of $\mu$ should not depend on $T,V$. Indeed while fitting to eq. \ref{eq:PnFinalFormula} we were always able to extract $\mu$ regardless of $T$ and $V$. But $T$ and $V$ are not fully independent in fitting parameters. This can be seen in Fig. \ref{fig:cmp4_chisq200} where we fix $\mu$ and plot chi-squared value as function of $T$ and $V$ (this figure is plotted for RHIC energy 200 GeV and LQCD data \cite{CA2016PRD}). From this figure it is clear that there is dependence T(V) which describes experimental data with the lowest chi-squared. However, if we plot chi-squared value along minimum chi-squared line $T=T(V)$ in Fig. \ref{fig:cmp4_chisq200} we will see different fit quality. In Fig. \ref{fig:chisq_valley200} we plot logarithm of chi-squared value along this line.

\begin{figure}[h]
	\includegraphics[width=\linewidth]{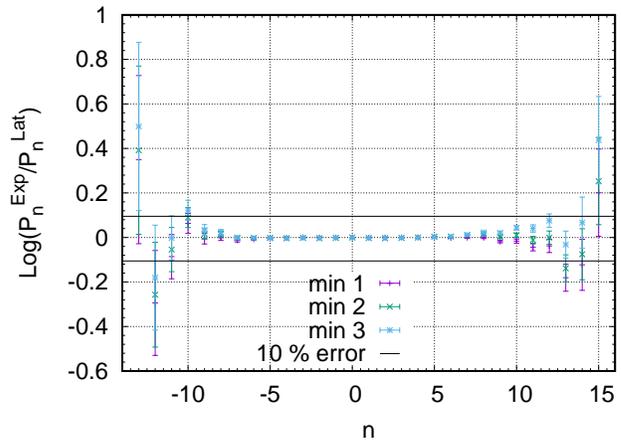}
	\caption{
		RHIC experimental multiplicity at 200 GeV and approximation with Grand Canonical Distribution with 5 set of parameters (T,V) from line of the best chi-squared T(V) on Fig. \ref{fig:cmp4_chisq200} and \ref{fig:chisq_valley200} for LQCD data \cite{CA2016PRD}, chemical potential is fixed to $\mu/T=0.16$.
	}
	\label{fig:fitQuality-s200}
\end{figure}
To show agreement between $P^{Lat}_n$ computed via eq.(8) and experimental
values of the multiplicity $P^{Exp}_n$ we depict in Fig. 5 the logarithm of 
their ratio for the lowest minimum of $\chi^2$.
One can see that agreement is within error bars for all values of $n$.

We repeated our analysis using state-of-the-art LQCD results for constants $f_k^{Lat}$ published in \cite{Fodor2017}. These results were obtained for improved staggered fermion action at physical quark masses. We extracted first two coefficients $f_k^{Lat}$ from Fig. 1 of this paper and computed the third coefficient using Cluster Expansion Model \cite{CEMPaper} expression (see eq. (5) in \cite{CEMPaper} ).
In Fig. \ref{fig:chisq_valley200} one can see for this set of LQCD data rather wide range of the lattice size $L_s$ values with low values of $\chi^2$ with only one (though 
rather wide) minimum.  It might be an indication that more coefficients $f_k$ 
allow us to fix parameters T and V more precisely. Comparison of $P^{Lat}_n/P^{Lat}_0$ and $P^{Exp}_n/P^{Exp}_0$ for other  experimental energies are shown in Appendix, Fig. \ref{fig:PnnFitAll}.

Next we show dependence of $\chi^2$ and parameters $T, V, \mu/T$ on RHIC energy. 
\begin{figure}[!h]
	\includegraphics[width=\linewidth]{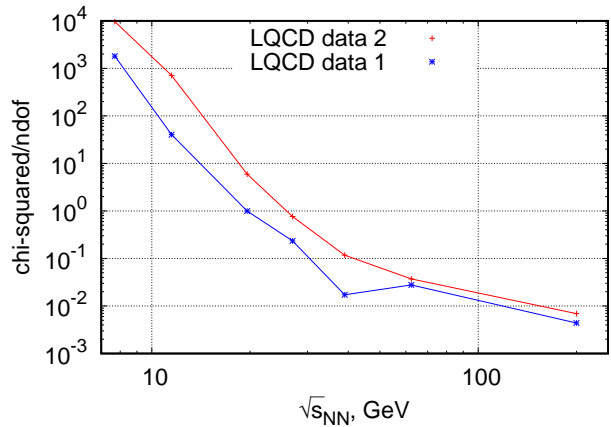}
	\caption{
		Chi-squared value of experimental data fit to Eq. \ref{eq:PnFinalFormula} with LQCD data, LQCD data 1 on figure stands for data \cite{CA2016PRD} and LQCD data 2 - for \cite{Fodor2017}.
	}
	\label{fig:PnChi2}
\end{figure}
In Fig. \ref{fig:PnChi2} we see $\chi^2$ is depicted. For energy higher 
than 19.6 GeV we find low $\chi^2$ while fits at low energies have large $\chi^2$
values.
\begin{figure}[!h]
	\includegraphics[width=\linewidth]{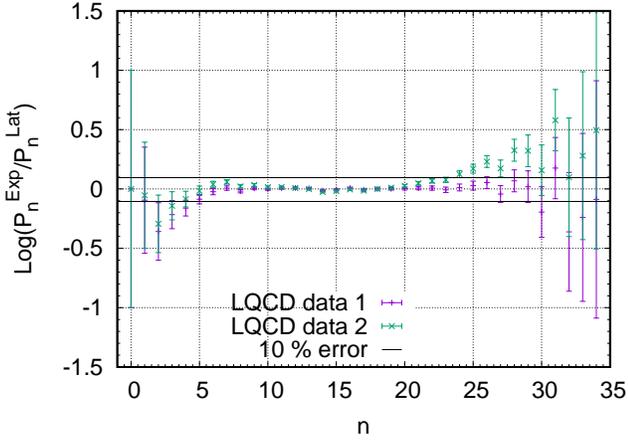}
	\caption{
		Comparison of fits of experimental data at $\sqrt{s_{NN}}$ = 7.7 GeV with different Lattice data, LQCD data 1 on figure stands for data \cite{CA2016PRD} and LQCD data 2 - for \cite{Fodor2017}.
	}
	\label{fig:LogPnFit77}
\end{figure}
From Fig. \ref{fig:LogPnFit77} one can see the systematic deviation of the fitting function from the experimental data at large $n$. This deviation gives rise to very large $\chi^2$. We thus find that fitting of experimental data at small energies and large $n$ requires probably modification of the fitting function for the quark number density, Eq. \ref{eq:numDens}. Also we admit the possibility that hadron matter created at small energies is not fully thermalized and Grand Canonical Distribution can be used only as a proxy.

\begin{figure}[!h]
	\includegraphics[width=\linewidth]{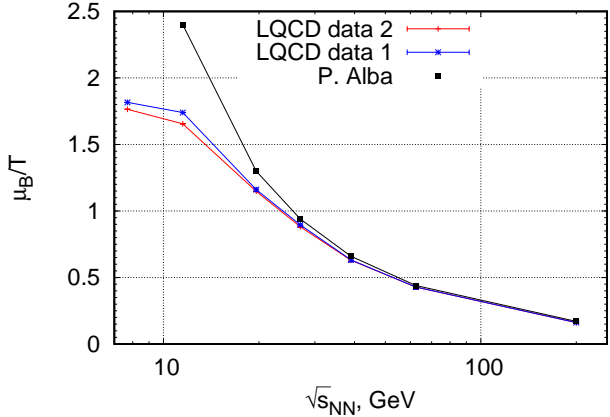}
	\caption{
		Extracted chemical potential of Grand Canonical Distributions of RHIC experimental data for different energies, P. Alba stands for \cite{Alba}, LQCD data 1 on figure stands for data \cite{CA2016PRD} and LQCD data 2 - for \cite{Fodor2017}.
	}
	\label{fig:ExpMu}
\end{figure}

In Fig \ref{fig:ExpMu} results for the baryon chemical potential $\mu/T$ are presented. 
They are in good agreement with estimations of Ref \cite{Alba}, especially at high 
energy.
For small energies we see that different Lattice data gives us slightly different results. This fact may contradict Eq. \ref{eq:PnFormula} because dependents of $P_n$ on $\mu$ and $T,V$ is factories so determination of $\mu$ should not depend on Lattice data, which have information about $T,V$. 
We think this disagreement may arise due to distortion of Grand Canonical Distribution. The magnitude of corresponding systematic error can be estimated from the discrepancy between different Lattice data on Fig. \ref{fig:ExpMu}.

\begin{figure}[!h]
	\includegraphics[width=\linewidth]{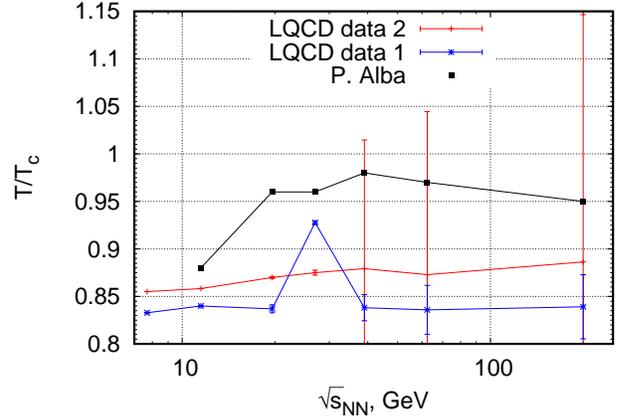}
	\caption{
		Extracted temperature of Grand Canonical Distributions of RHIC experimental data for different energies, P. Alba stands for \cite{Alba}, LQCD data 1 on figure stands for data \cite{CA2016PRD} ($T/T_c \approx 175\,\text{MeV}$) and LQCD data 2 - for \cite{Fodor2017} ($T/T_c \approx 155\,\text{MeV}$).
	}
	\label{fig:ExpT}
\end{figure}

Energy dependence of the fitted temperature $T$ is plotted in Fig. \ref{fig:ExpT}.
Its values are in agreement with findings of \cite{Alba} and indicate that $T$ is below
$T_c$ most likely.

\begin{figure}[!h]
	\includegraphics[width=\linewidth]{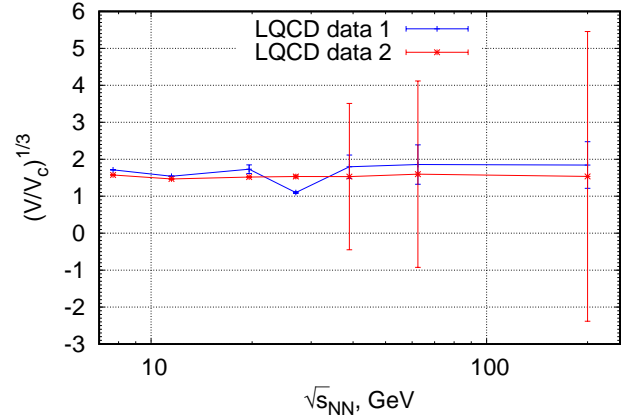}
	\caption{
		Extracted volume of Grand Canonical Distributions of RHIC experimental data for different energies, LQCD data 1 on figure stands for data \cite{CA2016PRD} ($V_c \approx (4.5\,\text{fm})^3$) and LQCD data 2 - for \cite{Fodor2017} ($V_c \approx (5.1 \,\text{fm})^3$).
	}
	\label{fig:ExpV}
\end{figure}

Finally, we present volume of fireball for different RHIC energies - Fig. \ref{fig:ExpV}.
As we understand it is the first estimation of volume of experimental fireball from Lattice QCD data. This value of the volume is consistent with Hanbury Brown and Twiss analysis of heavy ion collision experiments \cite{HICVolume}.

\section{Discussion and Conclusion}
In this paper we suggested a method of extracting freezout parameters from the multiplicity distributions measured in the heavy ion collisions. We fit STAR experiment net-proton multiplicity data used as an approximation of the net-baryon multiplicity with grand canonical distribution based on LQCD results. We show that there is a dependence of T(V) which allows describe multiplicity for net number $n<=7$. Multiplicity for higher net number is very important when one is aiming to define temperature and volume of fireball more precisely.
In both cases we obtained similar values for parameters $\mu/T, T/T_c, V/V_c$. This allows us to conclude that systematic errors due to lattice artifacts and large quark mass are not large and are partially encoded in the value of $T_c$ which is different at physical quark masses used in \cite{Fodor2017} and at large quark masses used in \cite{CA2016PRD}.
We shall note one difference in the output of our analysis for two sets of LQCD data. For data set 1 taken from \cite{CA2016PRD} we found three rather narrow minima of $\chi^2$ while for data set 2 taken from \cite{Fodor2017} only one but rather wide minimum was observed.

Two further conclusions can be drawn from our results. Our success (at least for high enough collision energies) in fitting the experimental multiplicity distributions with Grand Canonical Distributions obtained from first principle LQCD simulations implies that the hadronic matter in the fireball is in a thermalized state. This success also implies that the experimental net-proton multiplicities $P_n/P_0$ are good approximation of the net-baryon multiplicities.

As we noticed in the previous section we did not find low minima of $\chi^2$ for two lowest collision energies. We see three different sources potentially contributing to this failure. First, at low energies collision fragments can interact with fireball due to small velocity. Second, hadronic matter may be not fully thermalized at low energies. Third, difference between the net-proton distribution and the net-baryon one at low energies can be higher than at high energies.

The programming code for this paper was written with C++ programming language and can be found at github.com/boydad/PnFitWithTMuV. Fit was done with CERN Root Minuit2 library \cite{Minuit2}. For multi precision arithmetic MPFR C++ library \cite{MpfrCpp} was used, for Fast Fourier transform - Eigen library \cite{Eigen}.  

\section*{Appendix}
\begin{figure*}
	\centering
	\begin{minipage}{.48\linewidth}
		\centering
		{\includegraphics[width=\linewidth]{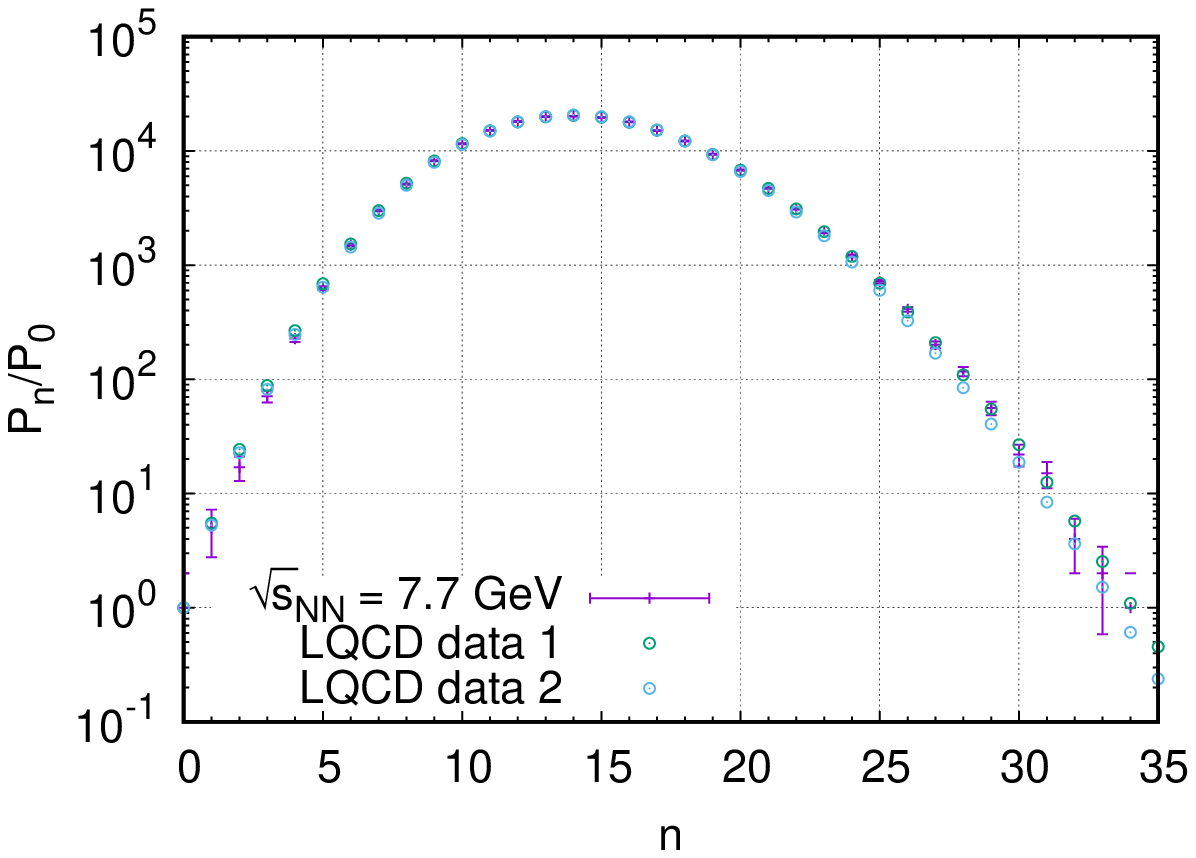}}
		
		{\includegraphics[width=\linewidth]{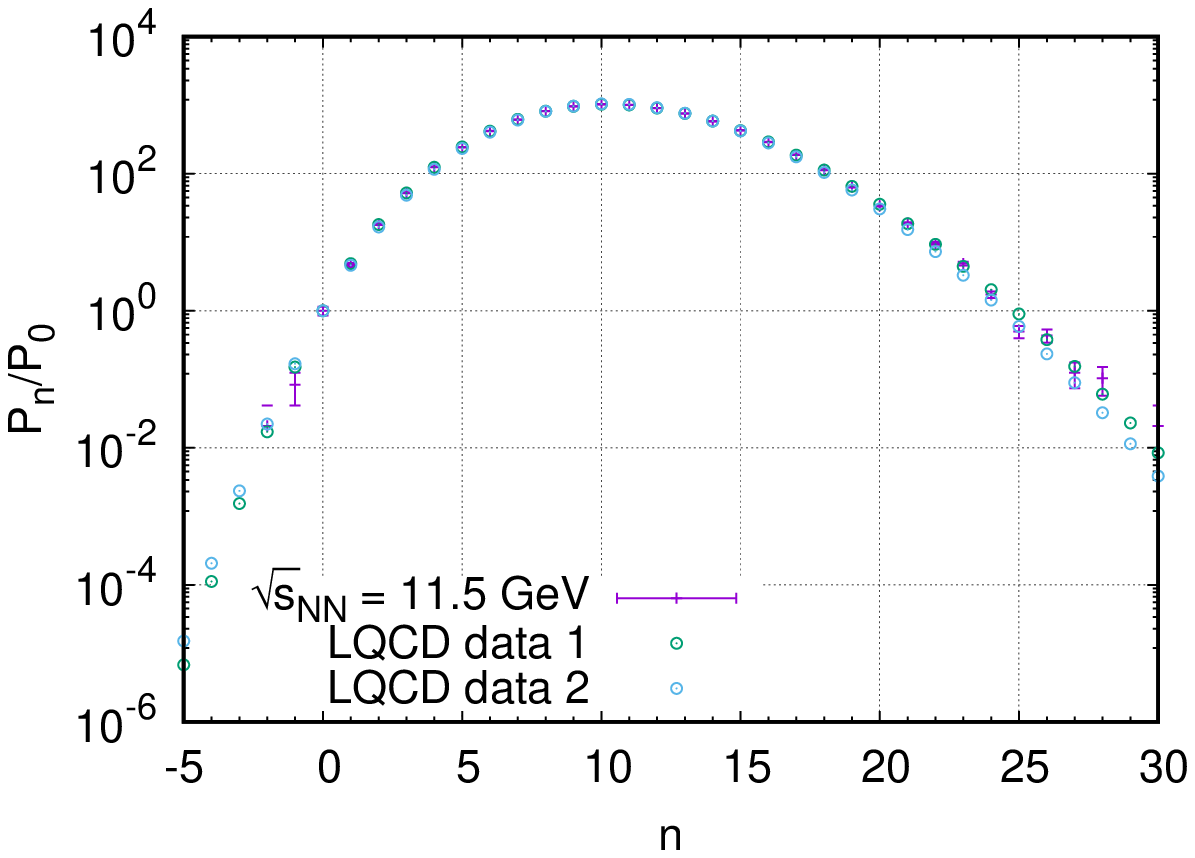}}
		
		{\includegraphics[width=\linewidth]{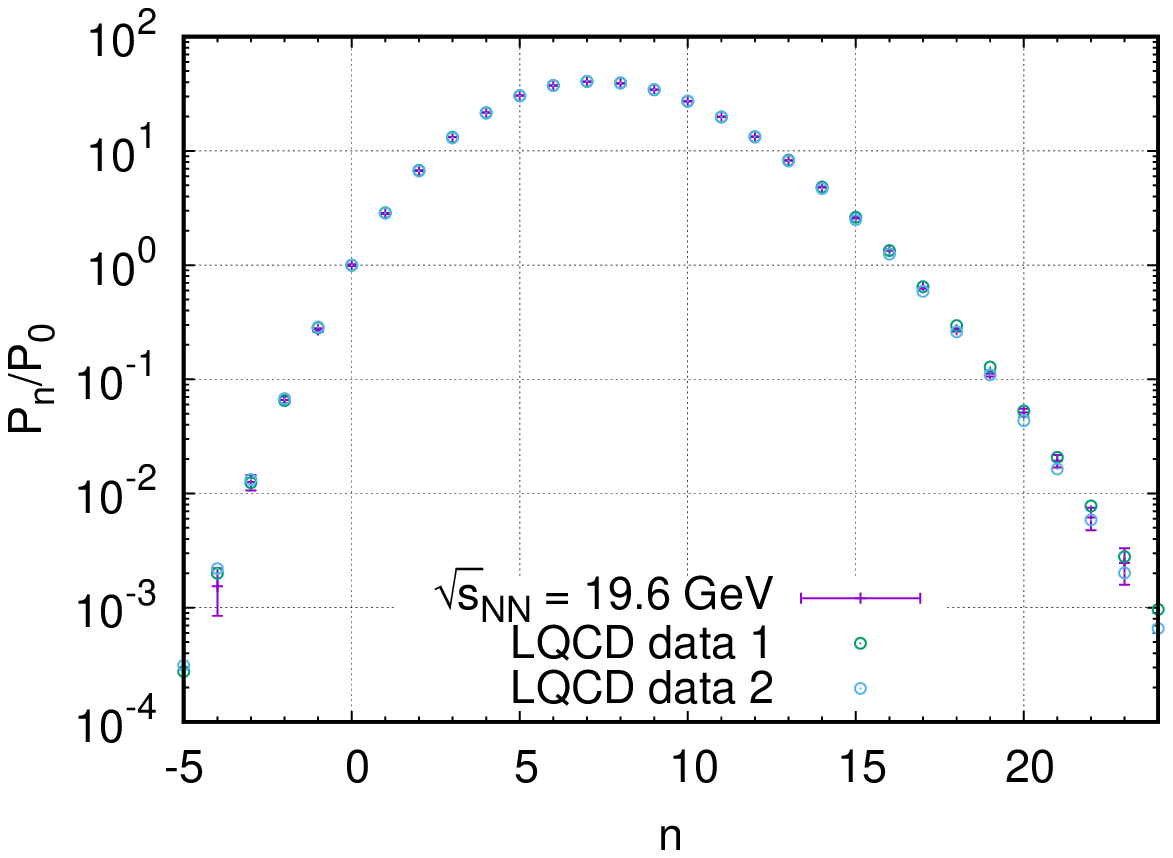}}

	\end{minipage}\quad
	\begin{minipage}{.48\linewidth}
		\centering
		{\includegraphics[width=\linewidth]{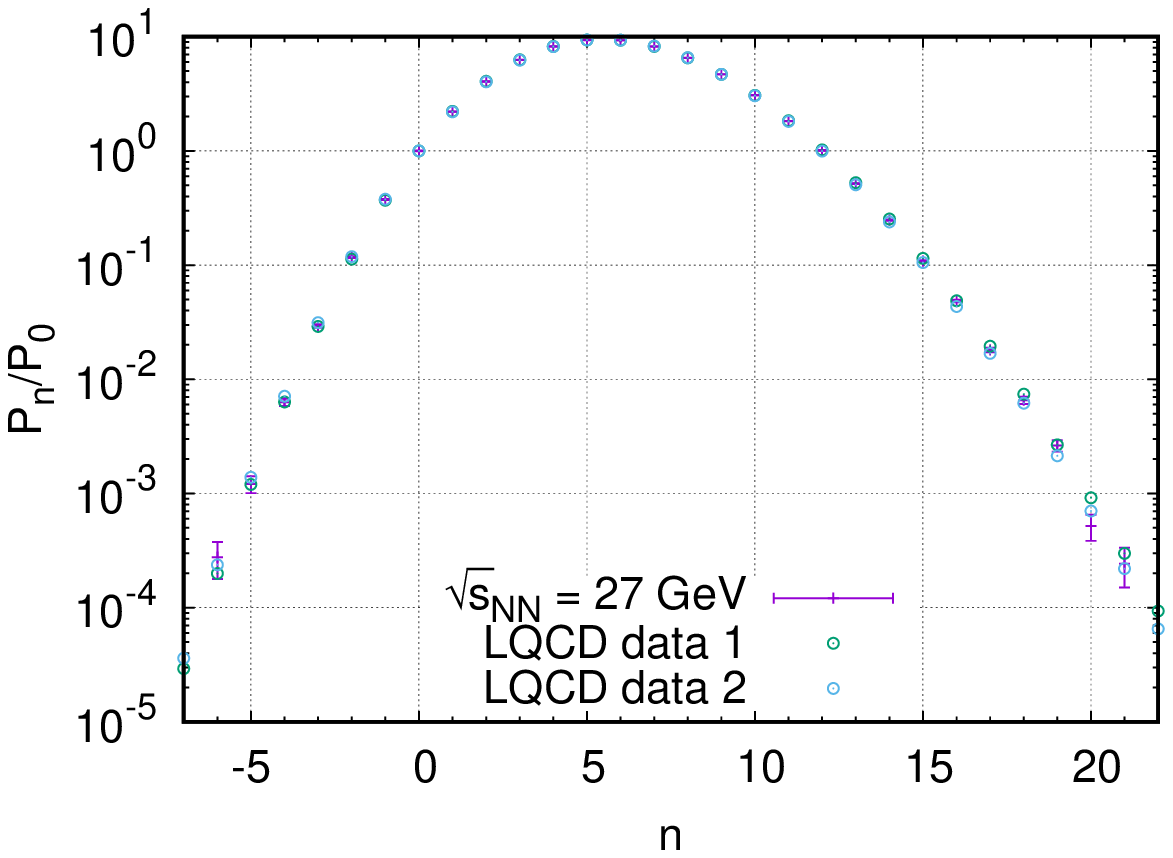}}
		
		{\includegraphics[width=\linewidth]{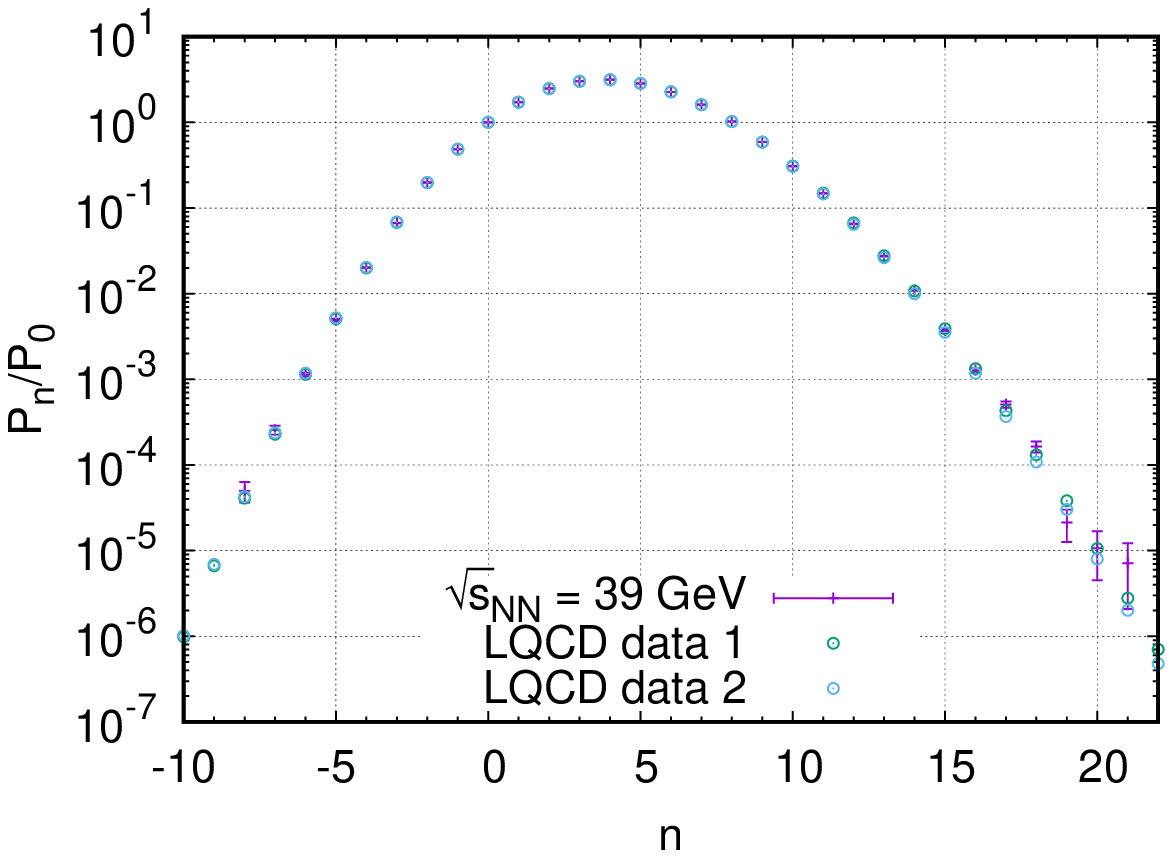}}
		
		{\includegraphics[width=\linewidth]{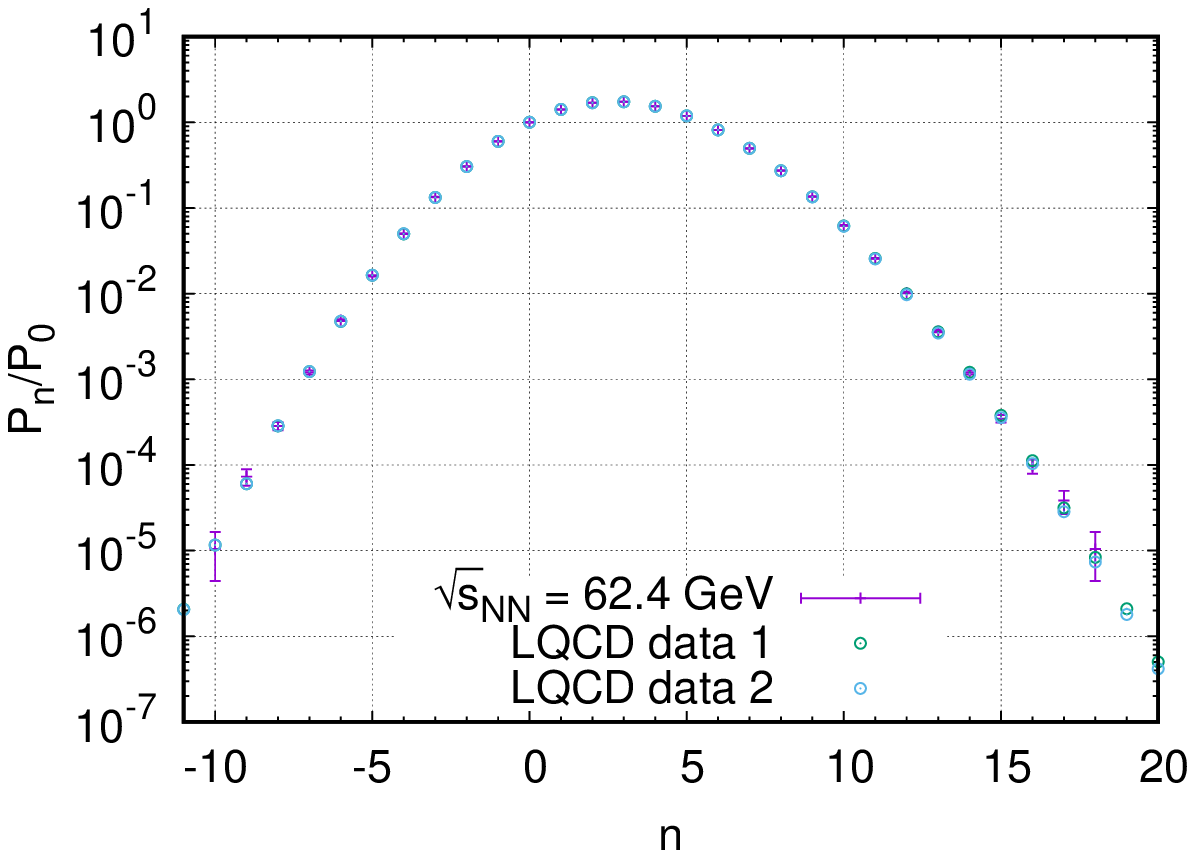}}
		
	\end{minipage}
	\caption{\label{fig:PnnFitAll}Fit experimental multiplicity to Canonical Distribution and different Lattice data with temperature, volume and chemical potential as parameters.}
\end{figure*}


\input{paper.bbl}
\end{document}

%% file: macroses.tex
\newcommand{\beq}{\begin{equation}}
\newcommand{\eeq}{\end{equation}}
\newcommand{\beqa}{\begin{eqnarray}}
\newcommand{\eeqa}{\end{eqnarray}}